\documentclass[preprint,aps]{revtex4}

\begin{document}
\draft
\title{Beating the PNS attack in practical
quantum cryptography}
\author{Xiang-Bin Wang\thanks{Email address: wang@qci.jst.go.jp}\\
IMAI Quantum Computation and Information Project,
ERATO, JST,
Daini Hongo White Bldg. 201, \\5-28-3, Hongo, Bunkyo,
Tokyo 133-0033, Japan}
\begin{abstract} 
 In practical quantum key distribution, weak coherent state is often used
and the channel transmittance can be very small therefore the protocol
could be totally insecure under the photon-number-splitting attack.
We propose an efficient method to verify 
the upper bound of the fraction of counts caused by multi-photon pluses
transmitted from Alice to Bob,
given whatever type of Eve's action. The protocol simply
uses two coherent states for the signal pulses and vacuum for decoy pulse.
 Our verified upper bound is sufficiently tight for QKD with very lossy 
channel,
in both asymptotic case and non-asymptotic case. The coherent states
with mean photon number from 0.2 to 0.5 can be used in practical 
quantum cryptography. We show that so far our protocol is the $only$ decoy-state
protocol
that really works for currently existing set-ups.    
\end{abstract}
\maketitle
Unlike the classical cryptography, quantum
key distribution(QKD)\cite{wies,gisin,bene} can help
two remote parties to set up the 
secure key by non-cloning theorem\cite{woot}.
Further, proofs for the unconditional security over
noisy channel have been given\cite{shor2,lo3,maye,ekert}.
The security of practical QKD with weak coherent states has also been shown\cite{inl,gllp}. 
However there are still some limitations for QKD in practice, especially over
long distance.
 In particular,
large loss of channel seems to be the main challenge to the long-distance
QKD with weak coherent states.
A dephased coherent state $|\mu e^{i\theta}\rangle$
is actually a mixed state of
\begin{eqnarray}
\rho_u=\frac{1}{2\pi}
\int_0^{2\pi}|\mu e^{i\theta}\rangle\langle\mu e^{i\theta}|{\rm d}\theta
=\sum_n P_n(\mu)|n\rangle\langle n|\label{coherent0}
\end{eqnarray}   
and 
$
P_n(\mu)=\frac{\mu^ne^{-\mu}}{n!}.
$
Here
$\mu$ is a non-negative number.
In practice, especially in doing long-distance QKD, the channel
transmittance $\eta$ can be rather small. If 
$\eta<(1-e^{-\mu}-\mu e^{-\mu})/\mu$, Eavesdropper (Eve) in principle
can have the full information of Bob's sifted key by the photon-number-splitting
(PNS) attack\cite{bra}:
Eve blocks all single-photon pulses and part of multi-photon pulses and 
separates each of the remained multi-photon pulses into two parts therefore each part contains at least one
photon. She keeps one part and sends the other part to Bob, through a 
lossless channel. 

If the channel is not so lossy, Alice and Bob can still
set-up the unconditionally secure final key with a key rate\cite{gllp} 
\begin{eqnarray}
r=1-\Delta-H(t)-(1-\Delta)H(t/(1-\Delta))
\label{dllp}
\end{eqnarray}
if we use a random classical CSS code\cite{shor2} to distill the final key\cite{gllp}. 
Here $t$ is the detected  flipping error rate, $\Delta$ is the fraction of tagged signals\cite{gllp},
i.e. the fraction for those counts in cases when
 Alice sends out a multi-photon pulse. 
The functional $H(x)=-x\log_2x-(1-x)\log_2(1-x)$.
From the above formula we see that verifying a tight
  bound for $\Delta$ is the first  important thing in QKD.

Originally, the PNS attack has been investigated where Alice and Bob 
monitor only how many non-vacuum signals arise, and how many errors 
happen. However, it was then shown\cite{kens1} that the simple-minded method
does not guarantee the final security. It is shown\cite{kens1}  that in a 
typical parameter regime nothing changes if one starts to monitor the 
photon number statistics as Eve can adapt her strategy to reshape the 
photon number distribution such that it becomes Poissonian again. 
A very important method with decoy states was then proposed by Hwang\cite{hwang}, where 
the {\it unconditional} verification of the multi-photon counting rate  is given.
 Hwang's decoy-state method can faithfully estimate the upper bound
of $\Delta$ through  decoy-pulses, given $whatever$ type of PNS attack.
(Remark: Decoy-state method is not the only solution to the issue. An 
alternative
method is to use strong refenence light\cite{srl}.)
 However, Hwang's initial protocol\cite{hwang} does not give
a sufficiently tight bound. 
 For example, in the case of $\mu=0.3$,
by Hwang's method,
the the optimized verified upper bound of $\Delta$ is $60.4\%$.
As it has been mentioned\cite{hwang,tot}, decoy-state method can be
combined with GLLP\cite{gllp}  to distill unconditionally secure final key.
With the value $\Delta=60.4\%$,  by eq(\ref{dllp}),
the key rate can be rather low in practice.
Following Hwang's work\cite{hwang}, decoy-state method was then studied by Lo and co-workers
\cite{tot,lo4}.
 They proposed their main protocol
: Try EVERY Poisson distribution of mixed states in Fock space, i.e.,
to test the counting rates of coherent
states $\{|\mu' e^{i\theta}\rangle\}$ with ALL possible values of $\mu'$ 
in one protocol.
In such a way the counting rates of each state $|n\rangle\langle n|$ can
be calculated therefore an exact value of $\Delta$ can be given.
However, such a protocol seems to be inefficient in practice, because it requires
infinite number of classes of different coherent states to work as the decoy states. 
Prior to this, an idea of using vacuum to test the dark count and using
$very$ weak coherent state as decoy state to verify the lower bound of single-photon transmittance 
had been shortly stated\cite{lo4}.  However, 
as it is shown in the Appendix, given a very lossy channel, the number of single-photon counts 
of all those decoy states is much less than the dark count. Even a {\em little bit} fluctuation of dark count
can totally destroy any meaningful estimation of single-photon counts.
To make a meaningful estimation, the statistical fluctuation of dark count must be pretty small
and this requires an unreasonablly large number of decoy-pulses which request more than 14 days to produce. 
In summary, {\bf so far no prior art result of decoy-state can really work in practice.} 
 
Here, we propose a new decoy-state
protocol which is the $only$ one that really works for currently
existing set-ups. The protocol uses only 3 different states 
and the verified bound values are sufficiently tight for long-distance QKD.
In the protocol, coherent states with average photon number $\mu$, $\mu'$ are used
for signal pulses and vacuum is used for the decoy pulse.
 Since both $\mu$ and $\mu'$ are in a reasonable range,
  pulses produced in both states 
can be used to distill the final key.
 That is to say, both of them can be
regarded as the {\it signal} states. 
 Our result {\it uniquely} shows that
 long-distance QKD with decoy states by a {\it real-world} 
protocol is possible, with a reasonable number of total pulses.

For simplicity, we denote those pulses produced in state  
$|\mu e^{i\theta}\rangle, |\mu' e^{i\theta}\rangle,|0\rangle$ as
class $Y_\mu, Y_{\mu'}$ and $Y_0$, respectively.
In the protocol $\theta$
is randomized. Also, Alice mixes the positions of all pulses therefore no one but Alice
knows which pulse belongs to which class in the protocol.
They observe the counting rates of each classes and then
verify the upper bounds of counts caused by multi-photon pulses
from class $Y_\mu,Y_{\mu'}$, respectively.
If these values are too large, they abandon 
 the protocol, otherwise they go on to 
do key distillations using pulses from each class of  $Y_\mu$
{\it and}  $Y_{\mu'}$ by GLLP\cite{gllp}.

We first define the {\it counting rate} of {\it any} state $\rho$:
 the probability that Bob's detector clicks whenever
a state $\rho$ is {\it sent out} by Alice.
We $disregard$ what state Bob may receive here. This {\it counting rate}
is called as the {\it yield} in other literatures\cite{hwang,tot}.
We denote the counting rate (yield) of vacuum, class $Y_0,Y_\mu,Y_{\mu'}$
by notations $s_0,S_\mu,S_{\mu'}$, respectively. These 3 parameters are observed in
the protocol itself: After all pulses are sent out, Bob announces which pulse has
caused a click which pulse has not caused a click. Since Alice knows which pulse
belongs to which class,  Alice can calculate the {\it counting rates} 
of each classes of pulses. Therefore we shall regard $s_0,S_\mu, S_{\mu'}$ 
as known parameters in protocol. The value $s_0$ is
counting rate at Bob's side when Alice sends vacuum pulses. We shall also
call $s_0$ as the vacuum count or dark count rate.

Their task is to verify
the upper bound of $\Delta$, the fraction of multi-photon counts among all counts caused
by pulses in  class $Y_\mu$ and also the upper bound of $\Delta'$, the fraction of 
multi-photon counts among all counts caused
by pulses in  class $Y_{\mu'}$.
We shall show how they can deduce the upper values of $\Delta,\Delta'$ from
the values of $\{s_0,S_\mu,S_{\mu'}\}$.
We shall focus on $\Delta$  first and latter obtain $\Delta'$ based on the knowledge of $\Delta$.
 
 For convenience, we $always$ assume 
\begin{eqnarray}
\mu'>\mu; \mu' e^{-\mu'} > \mu e^{-\mu} \label{condition}
\end{eqnarray} in this paper.
 A dephased coherent state $|\mu e^{i\theta}\rangle$ 
has the following convex form: 
\begin{eqnarray}
\rho_{\mu}= e^{-\mu}|0\rangle\langle0|+\mu e^{-\mu}|1\rangle\langle 1|
+c\rho_c\label{oo}
\end{eqnarray}
and $c=1-e^{-\mu}-\mu e^{-\mu}>0$, 
\begin{eqnarray}
\rho_c=\frac{1}{c}\sum_{n=2}^\infty P_n(\mu)|n\rangle\langle n|\label{coherent}.
\end{eqnarray}
Similarly, state  $|\mu' e^{i\theta}\rangle$ after dephasing is
\begin{eqnarray}
\rho_{\mu'}=e^{-\mu'}|0\rangle\langle0|+\mu' e^{-\mu'}|1\rangle\langle 1|
+c\frac{\mu'^2 e^{-\mu'}}{\mu^2 e^{-\mu}}\rho_c + d\rho_d\label{dd}
\end{eqnarray}
and $d=1-e^{-\mu'}-\mu' e^{-\mu'}-c\frac{\mu'^2 e^{-\mu'}}{u^2 e^{-\mu}} \ge 0$.
$\rho_d$ is a density operator. (We shall only use the fact that $d$ is non-negative and
$\rho_d$ $is$ a density operator.) In deriving the above convex form, we have used the fact
$P_n(\mu')/P_2(\mu')> P_n(\mu)/P_2(\mu)$ for all $n>2$, given the conditions of eq.(\ref{condition}).
With these convex forms of density operators,
it is equivalent to say that Alice sometimes sends nothing 
($|0\rangle\langle 0|$), sometimes sends $|1\rangle\langle 1|$,
sometimes sends $\rho_c$ and sometimes sends $\rho_d$,  
though Alice does not know which time she has sent out which one of these states. In each individual sending, 
she only knows which class the pulse belongs to.
We shall use notations $s_0,s_1,s_{c},S_\mu,S_{\mu'},s_d$ for the 
{\it counting rates} of state 
$|0\rangle\langle 0|, |1\rangle \langle 1|, \rho_c, \rho_\mu, \rho_{\mu'},\rho_d$,
respectively. 
 Given any state $\rho$, nobody but Alice can tell whether it is from
class $Y_\mu$ or $Y_{\mu'}$. Asymptotically, we have  
\begin{eqnarray}
s_\rho(\mu)= s_\rho(\mu')
\end{eqnarray} 
and $s_\rho(\mu),s_\rho(\mu')$ are {\it counting rates} 
for state $\rho$ from class $Y_\mu$ and class 
$Y_{\mu'}$, respectively. 

We shall use the safest assumption that 
Eve also controls the the detection
efficiency and dark count of Bob's detector. 
We only consider the overall transmittance 
  including
the channel, Bob's devices and detection efficiency.
 By eq.(\ref{oo}), we relate $\Delta$ with parameter $s_c$, the multi-photon
counting rate in class $Y_\mu$ by: 
\begin{eqnarray}
\Delta=c\frac{s_c}{S_{\mu}}.\label{new}
\end{eqnarray}
To verify the upper bound of  $\Delta$ for pulses from
class $Y_\mu$, we only need to 
verify the upper bound 
of $s_c$, the {\it counting rate} of mixed state $\rho_c$. 
The task is reduced to formulating $s_c$ by  $\{s_0,S_{\mu}, 
S_{\mu'}\}$, which are measured directly in the protocol itself.
The coherent state $\rho_{\mu'}$ 
is convexed by $\rho_c$ and other states. Given the condition
of eq.(\ref{condition}),  the probability of
$\rho_c$ in state $\rho_{\mu'}$ is larger than that in $\rho_{\mu}$. 
Using this fact we can make a preliminary estimation of $s_c$. 
From eq.(\ref{dd}) we immediately obtain 
\begin{eqnarray}
S_{\mu'}= e^{-\mu'}s_0 + \mu' e^{-\mu'}s_1 + c\frac{\mu'^2 e^{-\mu'}}
{\mu^2 e^{-\mu}}s_c +ds_d.\label{origin}
  \end{eqnarray}
$s_0$ is known, $s_1$ and $s_d$ are unknown, but they are never less than 0. Therefore
we have
\begin{eqnarray}
cs_c\le \frac{\mu^2e^{-\mu}}{\mu'^2e^{-\mu'}}\left(S_{\mu'}- e^{-\mu'}s_0-\mu' e^{-\mu'}s_1\right).
\label{origin8} 
\end{eqnarray} We  can obtain Hwang's main result\cite{hwang} by 
\begin{eqnarray}
cs_c\le \frac{\mu^2e^{-\mu}}{\mu'^2e^{-\mu'}}\left(S_{\mu'}- e^{-\mu'}s_0\right)
\le \frac{\mu^2e^{-\mu}}{\mu'^2e^{-\mu'}}S_{\mu'}\label{crude}
\end{eqnarray} 
Combining this equation with eq.(\ref{new})we have
\begin{eqnarray}
\Delta \le \frac{\mu^2e^{-\mu}S_{\mu'}}{\mu'^2e^{-\mu'}S_{\mu}}.
\end{eqnarray}
This is just eq.(12) in ref.\cite{hwang}. 
In the normal case that there is no
Eve's attack, Alice and Bob will find
$
S_{\mu'}/S_{\mu}=\frac{1-e^{-\eta\mu' }}{1-e^{-\eta\mu}}= \mu'/\mu
$ in there protocol therefore they they can verify 
$
 \Delta \le \frac{\mu e^{-\mu}}{\mu' e^{-\mu'}}
,$
which is just eq.(13) of Hwang's work\cite{hwang}.
Our derivation looks significantly simpler than that in ref.\cite{hwang}.

Having obtained the crude results above,
 we now show that the verification can be done
more sophisticatedly and one can further tighten the bound significantly.
In the inequality (\ref{crude}), we have dropped terms $s_1$ and $s_d$, since
we only have trivial knowledge about $s_1$ and $s_d$ there,
i.e., $s_1\ge 0$ and $s_d\ge 0$. Therefore, inequality(\ref{origin8}) has no advantage at that moment.
 However, after we have obtained
the crude upper bound of $s_c$, we can have a larger-than-0 lower
bound for $s_1$, provided that our crude upper bound for $\Delta$ given by eq.(\ref{crude}) is not too large.
 From eq.(\ref{oo}) we have
\begin{eqnarray}
e^{-\mu}s_0 + \mu e^{-\mu}s_1 + c s_c=S_{\mu}. \label{crude1}
\end{eqnarray}
 With the crude upper bound for $s_c$ given
by eq.(\ref{crude}), we have the non-trivial lower bound for $s_1$ now:
\begin{eqnarray}
s_1 \ge S_{\mu}-e^{-\mu}s_0 - c s_c
 > 0. \label{new1}
\end{eqnarray}
The updated
$s_1$ will in return further tighten the upper bound of $s_c$ by eq.(\ref{origin8}), 
and the tightened $s_c$ will again
update $s_1$ by eq.(\ref{new1}) and so on. After many iterations,  the final
values for $s_c$ and $s_1$ are given by the simultaneous constraints of 
of inequalities (\ref{new1}) and  (\ref{origin8}).
We have the following final bound after solving them:
\begin{eqnarray}
\Delta \le
\frac{\mu}{\mu'-\mu}\left(\frac{\mu e^{-\mu} S_{\mu'}}{\mu' e^{-\mu'} S_{\mu}}-1\right) 
+\frac{\mu e^{-\mu}s_0 }{ \mu' S_\mu }.\label{assym}
\end{eqnarray}
Here we have used eq.(\ref{new}). In the case of $s_0<<\eta$, 
if there is no Eve., $S_\mu'/S_\mu=\mu'/\mu$. 
Alice and Bob must be able to verify 
\begin{eqnarray}
\Delta = \left. \frac{\mu \left(e^{\mu'-\mu}-1\right)}{\mu'-\mu}\right|_{\mu'-\mu\rightarrow 0}=\mu
\end{eqnarray} 
in the protocol.
This is close to the real value of fraction of multi-photon counts: $1-e^{-\mu}$, given that $\eta<<1$. 
This shows that eq.(\ref{assym}) indeed gives a rather tight upper bound. 
In our derivation, all multi-photon counts from pulses in class $Y_\mu$ are due to only $one$
mixed state, $\rho_c$. Therefore we only need to calculate $one$ unknown parameter, $s_c$.
However, in Ref\cite{tot}, they have considered the contribution of each Fock state and there are
infinite number of unknown variables 
of $\{s_n\}$. Therefore they need $infinite$ number of different coherent states
in their main protocol\cite{tot} while we
only need three.

With the upper bound of  $s_c$ (or, $\Delta$), pulses from class $Y_\mu$ can be used for  
key distillation by GLLP\cite{gllp}.
On the other hand, given $s_c$, we can calculate the lower bound of $s_1$ through eq.(\ref{new1}).
Given $s_1$, we can also calculate the upper bound of $\Delta'$, the fraction
of multi-photon count among all counts caused by pulses from
class ${Y_{\mu'}}$. Explicitly,
\begin{eqnarray}
\Delta' \le 1- (
1-\Delta -\frac{e^{-\mu}s_0}{S_\mu})e^{\mu-\mu'}-\frac{e^{-\mu'}s_0}{S_{\mu'}}.
\end{eqnarray}
The values of  $\mu,\mu'$ should be chosen in a reasonable range, e.g., from 0.2 to 0.5.
To maximize the key rate, one need to consider the quantities of 
transmittance, quantum bit error rate(QBER) and vacuum counts jointly.
The optimization is not studied in this paper.

The results above are only for the asymptotic case. 
In practice, there are statistical fluctuations, i.e., Eve. has non-negligibly
small probability to
treat the pulses from different classes a little bit differently,
even though the pulses have the same state. 
It is $insecure$ if we simply use the asymptotic result in practice.
Our task remained is  to verify a tight upper bound
of $\Delta$ and the probability that the real value of $\Delta$ breaks the verified upper bound
is exponentially close to 0. 

The counting rate of any state $\rho$ in class $Y_{\mu'}$
now can be slightly different from the counting rate of the same state $\rho$ 
from another class, $Y_\mu$, with non-negligible
probability. We shall use the primed
notation for the counting rate for any state in class $Y_{\mu'}$ and the original notation
for the counting rate for any state in class $Y_\mu$. 
Explicitly, eq.(\ref{crude1},\ref{origin8})
are now converted to
\begin{eqnarray}
  \left\{ \begin{array}{l} 
e^{-\mu}s_0 + \mu e^{-\mu}s_1 + c s_c=S_{\mu},
 \\ cs'_{c}\le \frac{\mu^2e^{-\mu}}{\mu'^2e^{-\mu'}}\left(S_{\mu'}
- \mu' e^{-\mu'} s'_1
- e^{-\mu'}s'_0\right) .
  \end{array}
  \right. \label{couple}
 \end{eqnarray}
Setting $s_x' = (1-r_x)s_x$ for $x=1,c$  and $s'_0=(1+r_0)s_0$ we obtain
\begin{eqnarray}
\mu' e^\mu \left[(1-r_c)\frac{\mu'}{\mu}-1\right]\Delta \le \mu e^{\mu'}S_{\mu'}/S_\mu
-\mu'e^{\mu}+[(\mu'-\mu)s_0+r_1s_1+r_0s_0]/S_\mu.
\end{eqnarray}
From this we can see, if $\mu$ and $\mu'$ are too close, $\Delta$ can be very large. 
The important question here is now whether there are reasonable
values for $\mu',\mu$ so that our method has significant advantage to the
previous method\cite{hwang}. The answer is yes.

Given $N_1+N_2$ copies of state $\rho$,  suppose
the counting rate
for $N_1$ randomly chosen states is $s_{\rho}$ and the counting rate
for the remained states  is $s'_{\rho}$, the probability that $s_\rho-s'_\rho>\delta_\rho$
is
less than $\exp\left(-\frac{1}{4}{\delta_\rho}^2N_0/s_\rho\right)$
and $N_0 ={\rm Min}(N_1,N_2)$. Now we consider the difference of counting rates
for the same state from different classes, $Y_\mu$ and $Y_{\mu'}$.
 To make a faithful estimation 
for exponentially sure, we require 
${\delta_\rho}^2N_0/s_\rho =100$. This causes a relative fluctuation   
\begin{eqnarray}
r_\rho=\frac{\delta_\rho}{s_{\rho}}\le 10\sqrt{\frac{1}{s_{\rho}N_0}}\label{statis}.
\end{eqnarray} 
The probability of violation is less than $e^{-25}$.
 To formulate the relative fluctuation $r_1,r_c$
by $s_c$ and  $s_1$, we only need to check the number of pulses in $\rho_c$,
$|1\rangle\langle 1|$  in each classes in the protocol.
That is, using eq.(\ref{statis}),
we can replace $r_1,r_c$ in eq.(\ref{couple}) 
by $10e^{\mu/2}\sqrt{\frac{1}{\mu s_1N}}$, 
$10\sqrt{\frac{1}{c s_cN}}$, respectively
and $N$ is the number of pulses in class $Y_\mu$.
Since we assume the case where vacuum-counting rate is much less than
the counting rate of state $\rho_\mu$, we omit the effect of fluctuation
in vacuum counting, i.e., we set $r_0=0$.
 With these inputs, eq.(\ref{couple}) can now be solved
numerically.
 The results are listed in the following table. 
 From this table we can
see that good values of $\mu,\mu'$ indeed exist and our verified
upper bounds are sufficiently tight to make QKD over very lossy channel.
Note that so far this is the $only$ non-asymptotic result among all existing works on decoy-state. 
From the table we can see that our non-asymptotic values are less than Hwang$'$s asymptotic
values already. Our verified values are rather close to the true values. We have assumed the vacuum count rate
$s_0=10^{-6}$ in the calculation. If $s_0$ is smaller, our results will be even better. 
Actually, the value of $s_0$ (dark count) can be even
lower than the assumed value here\cite{gobby,tomita}. 
\begin{table}
\caption{The verified upper bound of
the fraction of tagged pulses in QKD.
$\Delta_H$ is the result from Hwang's method. $\Delta_R$ is the true value of the fraction of
multi-photon counts in case there is no Eve.
 $\Delta_H$ and $\Delta_R$ do not change with channel transmittance.
$\Delta_{W1}$ is bound for pulses
in class $Y_\mu$, given that $\eta=10^{-3}$.
 $\Delta_{W2}$ and $\Delta'_{W2}$  are bound values
for the pulses in class $Y_\mu,Y_{\mu'}$ respectively,
given that $\eta=10^{-4}$. 
We assume $s_0=10^{-6}$. The number of pulses is
 $ 10^{10}$  in class $Y_\mu,Y_\mu'$ in calculating $\Delta_{W1}$
and $8\times 10^{10}$ in calculating $\Delta_{W2},\Delta'_{W2}$. 
$4\times 10^{9}$ 
vacuum pulses
is sufficient for class $Y_0$. The bound values will change by less than
0.01 if the value of $s_0$ is 1.5 times larger. The numbers inside
brackets are chosen values for $\mu'$. 
For example, in the column of $\mu=0.25$,
data $30.9\%(0.41)$ means, if we choose $\mu=0.25,\mu'=0.41$, we can
verify $\Delta\le 30.9\%$ for class $Y_\mu$. }
\begin{tabular}{c|c|c|c|c}
$\mu$ & 0.2 & 0.25 &0.3 & 0.35  
\\ 
\hline 
$\Delta_H$ & 44.5\% &52.9\%&  60.4\% & 67.0\%\\ \hline
$\Delta_{R}$ & 18.3\% & 22.2\% & 25.9\%
& 29.5\%\\
\hline
$\Delta_{W1}$ & 23.4\%(0.34)& 28.9\%(0.38)& 34.4\%(0.43)
& 39.9\%(0.45)\\
\hline
$\Delta_{W2}$ & 25.6\%(0.39) & 30.9\%(0.41) & 36.2\%(0.45
)& 41.5\%(0.47)
\end{tabular}
\begin{tabular}{c|c|c|c|c}
$\mu'$ & 0.39 & 0.41 &0.45 & 0.47\\ \hline
$\Delta_H$ & 71.8\% & 74.0\% & 78.0\% & 79.8\%
\\ \hline
$\Delta_{R}$ & 32.3\% & 33.7\% & 36.2\%
& 37.5\% \\ \hline
$\Delta'_{W2}$ & 40.1\% & 42.2\% & 45.8\% & 48.6
\end{tabular}
\end{table}
In the real set-up given by Gobby et al\cite{gobby}, the light loses a half
over every 15km, the devices and detection loss is $4.5\%$ and
$s_0\le 8.5\times 10^{-7}$. 
Given these parameters, we believe that our protocol
works over a distance longer than 120km with 
 with $\mu=0.3,\mu'=0.45$ and {\em a reasonable number} of total pulses.

In conclusion, following the work by Hwang\cite{hwang}, we have proposed an efficient and feasible decoy-state 
method to
do QKD over very lossy channel. The main protocol in Ref.\cite{tot} is impractical
because it depends on infinite number of pulses. The idea stated in Ref\cite{lo4} doesn't
work in practice either because it has implicitly assumed an unreasonablly large numbe of
pulses which require more than 14 days to produce, by the currently existing technology.
 Our protocol is the $only$ decoy-state protocol which 
really works with currently existing set-ups. The method of this paper
can be further developed\cite{wang2}.

{\em Note added:} After the earlier versions of this work (quant-ph/0410075) had been presente\cite{wang0}, H. K. Lo et al
also presented their previously announced results in the arXiv (quant-ph/0411004)\cite{lolo}. They claim that
they have {\em for the first time} made the decoy-state method efficiently useful in practice. We question their 
claim as we have shown that none of their previously announced protocol or idea really works in practice.
If, in Ref\cite{lolo}, their claim is actually based on 
something different from their previously announced results, since Ref\cite{lolo} itself
is presented later than our work\cite{wang0}, then at least the phrase ``for the first time'' is inappropriate in their claim.
To my understanding, Ref\cite{lolo} itself does not contain anything new, it is an extended version
of their previously announced results. Therefore, not only the phrase ``for the first time'' in their claim is inappropriate,
but also their whole claim is inappropriate. 
\section*{Appendix:}
In this Appendix, we give detailed demonstration that the shortly stated idea in Ref\cite{lo4} actually doesn't work.
It is stated\cite{lo4}:``On one hand, by using a vacuum as decoy state, 
Alice and Bob can verify the so called dark count rates of their detectors. On the other
hand, by using a very weak coherent pulse as decoy state, Alice and Bob can easily lower bound
the yield (channel transmittance) of single-photon pulses.'' 
This is to say, there are two sets of
decoy pulses: Set $Y_0$ contains $M$ vacuum pulses $|0\rangle\langle 0|$ and
set $Y_v$ contains $N$ pulses of very weak coherent state $|\mu_v\rangle\langle \mu_v|$. 
They can only observe the total counts of set $Y_0$ and the total counts of set $Y_v$.
 By that idea\cite{lo4}, to verify a meaningful lower bound of
$s_1$, the value $\mu_v$  must be less than channel transmittance $\eta$.
For clarity, we assume zero dark count first. They can only observe the total counts 
of pulses in set $Y_v$.
In the normal case when there is no Eve,  $N$ decoy pulses in class $Y_v$ will cause $N(1-e^{-\eta\mu_v})$ counts.
For the security, one has no other choice but to assume the
worst case that all multi-photon pulses have caused a count. Therefore the lower bound
of single-photon counts is $N[1-e^{-\eta \mu_v}-(1-e^{-\mu_v}-\mu_v e^{-\mu_v})]=N(\eta\mu_v-\mu_v^2/2)$.
The lower bound value for $s_1$ is verified by  $s_1\ge \frac{N(\eta\mu_v-\mu_v^2/2)}{N\mu_ve^{-\mu_v}}\approx \eta -\mu_v/2$.
Therefore one has to request $\mu_v\le\eta$ here if one wants to verify 
$s_1\ge \eta/2$.
Now we consider the effect caused by dark counts.
Suppose, after observed the counts of pulses in set $Y_0$, they find that the dark count rate, $s_0=10^{-6}$
for set $Y_0$. Note that the dark count rate for set $Y_0$ and the dark count rate for set $Y_v$ can be a little bit different due to the
stastical fluctuation.    
Given $N$ pulses of state $|\mu_v\rangle$, there are $Ne^{-\mu_v}$ vacuum pulses
and $N(1-e^{-\mu_v})$ non-vacuum pulses. Alice does not know which
pulse is vacuum which pulse is non-vacuum. They can $only$ observe
the number of total counts ($n_t$) caused by $N$ decoy pulses in set $Y_v$, which is
the summation of dark counts, $n_0$, the number of single-photon counts $n_1$ and the number of
multi-photon counts, $n_m$, of those $N$ decoy pulses in set $Y_v$.
 After observed the number of total counts $n_t$, they try to estimates
$n_1$ by the formula $n_t=n_0+n_1+n_m$, with $n_0=Ns_0' e^{-\mu_v}$ and 
the worst-case assumption of $n_m= N(1-e^{-\mu_v}-\mu_v e^{-\mu_v})$.
The value $s_0'$ is the dark count rate for set $Y_v$ and the value $s_0'$
 is never known $exactly$. They only know the approximate value, $s_0'\approx s_0 =10^{-6}$. 
Consider the case $\eta=10^{-4}$. The expected value of $n_1+n_m$  is 
$N(1-e^{-\eta\mu})\le 10^{-8}N$. Meanwhile, the expected number of dark counts is around $ 10^{-6}N$.
Since the expected number of dark counts there is much larger than the expected number of $n_1+n_m$,
 {\em a little bit} fluctuation of dark counts will totally destroy the estimation
of the value $n_1+n_m$ therefore totally destroy the estimation of $n_1$.
 To make a 
faithful estimation, we request the fluctuation of dark count to be much less than 
$10^{-8}N$, e.g., in the magnitude order of $10^{-9}N$. This is to say,
 one must make sure that the relative fluctuation of dark counts
is less than
$0.1\%$, with a probability {\em exponentially} close to 1 (say, $1-e^{-25}$). This requires $N$ larger than $10^{14}$.
The system repetition rate is normally less than $8\times 10^7$ in practice. Producing $10^{14}$ decoy pulses needs more than
14 days.      
  
\acknowledgments
I am
 grateful to Prof. H. Imai for his long-term support. 
I thank Toshiyuki Shimono for his kindly help in the numerical calculation.

\end{document}